\documentclass[aip,cha,reprint,groupedaddress,showpacs,showkeys,floatfix]{revtex4-1}
\usepackage{graphicx, color}
\usepackage{amsmath}
\usepackage{amsfonts}
\usepackage{pgf}
\usepackage[breaklinks=true,colorlinks=true,linkcolor=blue,urlcolor=blue,citecolor=blue]{hyperref}
\usepackage[normalem]{ulem}

\newcommand{\qed}{\nobreak\ifvmode\relax\else
      \ifdim\lastskip<1.5em \hskip-\lastskip\hskip1.5em plus0em minus0.5em \fi \nobreak\vrule height0.75em width0.5em depth0.25em\fi}

\newcommand{\A}[1]{$\mathbf{A_{#1}}$}
\newcommand{\B}[1]{$\mathbf{B_{#1}}$}

\begin{document}

\title{Riddled Basins of Attraction in Systems Exhibiting Extreme Events}

\author{Arindam Saha}
\email{arindam.saha@uni-oldenburg.de}

\author{Ulrike Feudel}
\email{ulrike.feudel@uni-oldenburg.de}

\affiliation{Theoretical Physics/Complex Systems, ICBM, University of Oldenburg, 26129 Oldenburg, Germany}



\begin{abstract}
Using a system of two FitzHugh-Nagumo units, we demonstrate the occurrence of riddled basins of attraction in delay-coupled systems as the coupling between the units is increased. We characterize the riddled basin using the uncertainty exponent which is a measure of the dimensions of the basin boundary. Additionally, we show that the phase space can be partitioned into pure and mixed regions, where initial conditions in the pure regions certainly avoid the generation of extreme events while initial conditions in the mixed region may or may not exhibit such events. This implies, that any tiny perturbation of initial conditions in the mixed region could yield the emergence of extreme events because the latter state possesses a riddled basin of attraction.
\end{abstract}

\maketitle

\begin{quotation}
Extreme events are rare, recurrent, irregular events which have a large impact on the system. Examples of such events include natural disasters, financial crises, harmful algal blooms and rogue waves. Due to the huge and potentially disastrous consequences that such events might have, it is crucial to understand the mechanisms and initial configurations of the systems leading to the occurrence of such events. In a recent study, we showed that extreme events can emerge in a system of delay-coupled relaxation oscillators. Here we focus on the analysis of multistability, i.e.\ the coexistence of several states for a given set of parameters. This implies that it depends crucially on the initial condition which of those stable states will be realized in the long-term limit. We find for that system of delay-coupled oscillators that the basins of attraction, i.e.\ the set of initial conditions converging to one particular attractor, become progressively complex as we approach the parameter region in which extreme events occur. Our main finding is that the phase space of the system which contains all possible initial configurations of the system can be partitioned into `pure' regions where extreme events certainly do not occur and `mixed' regions where extreme events may or may not occur. Furthermore, points in the mixed region are scattered in a way that very tiny perturbations can change the dynamics from the one which exhibits extreme events to the one which does not. In particular, our analysis indicates that specifically the basin of attraction of the existence of extreme events possesses a riddled structure.
\end{quotation}

\section{Introduction} 
\label{sec:Intro}

Recurring events which have a large impact on the system and are infrequent and irregular are known in the literature as extreme events.~\cite{ansmann2013extreme, karnatak2014route}. Due to their occurrence in a large class of physical systems~\cite{RevModPhys.73.1067, Feigenbaum2001, dobson2007, bunde2002science}, a large body of research has been  devoted to understanding such events in specific systems like rogue waves in oceans~\cite{akhmediev2009waves, chabchoub2011rogue, chabchoub2012super} and coupled laser systems~\cite{bonatto2011deterministic, akhmediev2016roadmap, pisarchik2011rogue, dal2013extreme}, harmful algal blooms in marine ecosystems~\cite{bialonski2015data, bialonski2016phytoplankton}, epileptic seizures in the brain~\cite{lehnertz2008epilepsy, Lehnertz2006} and adverse weather conditions like floods, droughts and cyclones. Additionally, studies using theoretical models have shown that extreme events can be generated via various mechanisms including incoherent background of interacting waves~\cite{kim2003statistics}, noise-induced attractor hopping~\cite{reinoso2013extreme, zamora2013rogue}, pulse-coupled small world networks~\cite{rothkegel2014irregular}, inhomogeneous networks of oscillators~\cite{ansmann2013extreme, karnatak2014route} and delay coupled relaxation oscillators~\cite{PhysRevE.95.062219}. For some of those systems, parameter regions have been identified in which several attractors can coexist. Particularly interesting is the coexistence of attractors containing extreme events and attractors exhibiting only regular motion. In those cases, it depends crucially on the initial conditions whether extreme events would occur. The structure of basins of attraction is essential for assessing the risk of the emergence of extreme events, but this question has rarely been addressed in literature.

For several decades, significant research has been done in the field of basin structures and their boundaries in general dynamical systems. Some of the interesting basin structures include fractal basins~\cite{VANDERMEER2001265, grebogi2012chaos, MCDONALD1985125}, Wada basins~\cite{NUSSE1996242, doi:10.1142/S0218127496000035}, intermingled basins~\cite{SOMMERER1996243,PhysRevE.54.2489, PhysRevE.52.R3313} and riddled basins~\cite{doi:10.1142/S0218127492000446, PhysRevLett.73.3528, OTT1994384}. Various studies have analyzed the specific conditions which lead to the emergence of each of these basin types~\cite{PhysRevLett.50.935, KENNEDY1991213, ott1994blowout, PhysRevLett.71.4134}.For the study presented here riddled basins are of particular interest. A basin is said to be riddled, if any arbitrary neighborhood of every point of the basin contains points from another basin of attraction~\cite{ASHWIN1994126, 0951-7715-9-3-006}. An important consequence of this property is that, if the basin of any attractor is riddled, an arbitrarily small perturbation in any initial condition from the basin of attraction of this particular attractor, can make the system converge to another attractor. It is known that riddled basins are often formed in systems with certain symmetries which manifest themselves as invariant manifolds of the system~\cite{CAZELLES2001301}. In previous studies, riddled basins have been found in a variety of systems including simple maps~\cite{PhysRevE.57.2713, PhysRevE.56.6393}, electronic circuits~\cite{0305-4470-28-3-001} and instantaneously coupled chaotic oscillators~\cite{PhysRevLett.73.3528, doi:10.1063/1.4954022}. Here we present the occurrence of riddled basins in delay-coupled relaxation oscillators. This very complicated basin structure leads to an extremely high sensitivity of the system with respect to perturbations. The latter property can have a strong impact on the system's dynamics since the attractor possessing the riddled basin is the one containing extreme events.

In this paper, we present a system of delay-coupled FitzHugh Nagumo (FHN) oscillators which have been recently shown to exhibit extreme events~\cite{PhysRevE.95.062219} and investigate the emergence of various types of multistability, i.e.\ the coexistence of different attractors and their respective basins of attraction. In particular, we show that this system exhibits riddled basins of attraction in the parameter regime where extreme events are observed. After introducing the model in Sec.~\ref{sec:Model}, we discuss various regimes of multistability in Sec.~\ref{sec:Multistability}. We show that the basins of attraction become more and more complex as the coupling strength is increased. In particular, we identify a riddled basin of attraction belonging to an attractor exhibiting extreme events in Sec.~\ref{sec:Characteristics} by extending the concept of final state sensitivity to infinite-dimensional dynamical systems and providing further evidence for the riddled structure by classifying points as interior or boundary points. We underline the consequences of a riddled basin structure in a system exhibiting extreme events in the conclusions (Sec.~\ref{sec:Conclusions}).

\section{The Model}
\label{sec:Model}

We consider a pair of FHN units $(i=1,2)$, which are coupled to each other using two time delayed diffusive couplings. If the coupling strengths of the system are given by $M_1$ and $M_2$; and the respective time delays are given by $\tau_1$ and $\tau_2$, then the dynamical equations governing the system are given as,
\begin{equation}
\begin{aligned}
  \dot{x}_i &= x_i(a-x_i)(x_i-1)-y_i + \sum_{k=1,2} M_k (x_j^{(\tau_k)}-x_i) \\
\dot{y}_i &= bx_i-cy_i + \sum_{k=1,2 }M_k (y_j^{(\tau_k)}-y_i),
  \label{eq:Definition}
\end{aligned}
\end{equation}
where $x_j^{(\tau_k)}=x_j \left( t-\tau_k \right)$, $y_j^{(\tau_k)}=y_j \left( t-\tau_k \right)$ and $i \neq j$. The two FHN units possess identical parameters $a$, $b$ and $c$. For our investigations, we fix them at $a=-0.025$, $b=0.00652$ and $c=0.02$. These parameter values correspond to the regime where in absence of coupling, each FHN unit executes oscillatory behavior in the long term.

\begin{figure}
  \includegraphics[width=0.95\linewidth]{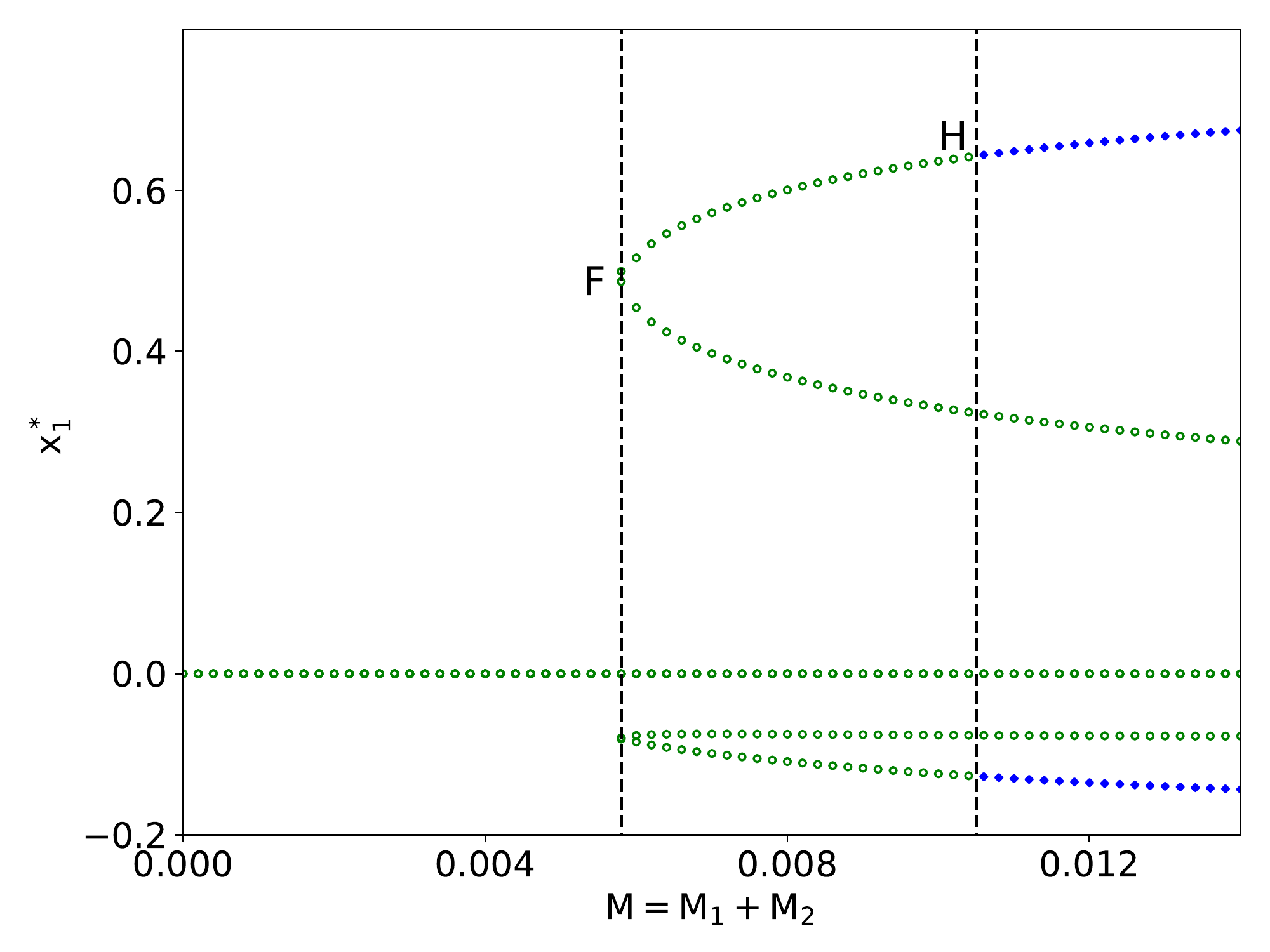}
\caption{Bifurcation diagram showing the position of the fixed points for varying coupling strength $M$. The green hollow circles represent unstable fixed points and the blue solid diamonds represent their stable counterparts. The points of fold and Hopf bifurcations are marked by `F' and `H' respectively.}
\label{fig:FixedPoints}
\end{figure}

Having identical internal parameters for FHN units implies the existence of an invariant manifold defined by $x_1^{(\tau)}=x_2^{(\tau)}$; $y_1^{(\tau)}=y_2^{(\tau)}$ for all $\tau \in \left[ 0, \text{max}\left\{ \tau_k \right\} \right]$. This manifold corresponds to the complete synchrony of the two units and partitions the phase space of the system in two symmetric halves. Note that this symmetry is particularly reflected in the position of the fixed points of this system in phase space. Hence, if $\left( x_1^*, y_1^*, x_2^*, y_2^* \right)$ is a fixed point of the system outside the synchronization manifold, then $\left( x_2^*, y_2^*, x_1^*, y_1^* \right)$ is also a fixed point with the same stability.

A useful way to exploit the symmetry of the system is by transforming Eq.~\ref{eq:Definition} to new coordinates, $X_{1,2}=\frac{x_1 \pm x_2}{2}$ and $Y_{1,2}=\frac{y_1 \pm y_2}{2}$. In these coordinates, $\left( X_1, Y_1 \right)$ denote the position of the projection of a general point $\left( X_1, Y_1, X_2, Y_2 \right)$ on the synchronization manifold and $\left( X_2, Y_2 \right)$ represents the separation between the point and the synchronization manifold. Moreover, if we define $X_j^{(\tau_k)}=X_j \left( t-\tau_k \right)$, any point on the synchronization manifold can be represented by $X_2^{(\tau)}=Y_2^{(\tau)}=0$ for all $\tau \in \left[ 0, \text{max}\left\{ \tau_k \right\} \right]$. Here, we use the transformed co-ordinates in figures to distinguish between attractors located on the synchronization manifold and the ones outside the synchronization manifold.

Moreover we note that, due to the form of coupling used in Eq.~\ref{eq:Definition}, it can be shown that the position of the fixed points is dependent neither on the time delays nor on the individual coupling strengths. Instead, it only depends on the sum of coupling strengths, $M=M_1+M_2$. Explicit computation yields several fixed points (See Fig.~\ref{fig:FixedPoints}) among which the origin is the only one present for the whole interval of coupling strength, even for zero coupling. As the coupling strength increases from zero, a pair of unstable fixed points appear on either side of the synchronization manifold via a fold bifurcation \textbf{F}. Thereafter, one of the fixed points on each side of the manifold stabilizes through a reverse Hopf bifurcation \textbf{H}.  Since we are mainly interested in the parameter regions in which multistability occurs, we focus on that interval of coupling strength in which two stable fixed points outside the synchronization manifold exist. However, we also consider the transition in which those two fixed points become stable. The dynamical properties of the system with no stable fixed points is studied in detail in our previous work~\cite{PhysRevE.95.062219}.

\section{Multistability and the Structure of Basins of Attraction}
\label{sec:Multistability}

According to our aim, we analyze different regimes of multistability in the system described in Eq.~\ref{eq:Definition} while the coupling parameters are varied. For the sake of simplicity, we keep two of the coupling parameters fixed at $M_1=0.01$ and $\tau_1=80$ throughout the article and discuss the changes in dynamics as parameters $M_2$ and $\tau_2$ are varied. To that end, we first fix $\tau_2=65$ and vary $M_2$. The effects of varying $\tau_2$ with fixed $M_2$ will be discussed briefly later in the article. The numerical simulations presented in this article were performed using the Python package JITCDDE~\cite{JITCDDE} which integrates systems of delay-differential equations using a modified form~\cite{SHAMPINE2001441} of the Bogacki-Shampine Runge-Kutta method.

Varying $M_2$ leads to various regimes, each of which is characterized by its own set of coexisting attractors and their corresponding basins of attraction. Due to the large number of attractors encountered during the parameter sweep, we index the $i^{\text{th}}$ attractor encountered by the symbol \A{i} and its corresponding basin of attraction by \B{i}. Also note that, since the phase space for a continuous time-delayed system is infinite dimensional, it is not possible to show the entire phase space in a diagram. We therefore show only slices of the phase space where the initial history of the system is identical to the initial conditions of the system: $x_i(\tau)=x_i(0)$ and $y_i(\tau)=y_i(0)$ for $-\text{max}\left\{ \tau_1, \tau_2 \right\} \le \tau < 0$. This brings down the dimensions of the slice to four, which is further reduced to two by choosing $y_1(0)=y_2(0)=0.01$. The structure of the basins of attraction in other slices of the phase space will be discussed in a later section of the article.

\begin{figure}
  \includegraphics[width=\linewidth]{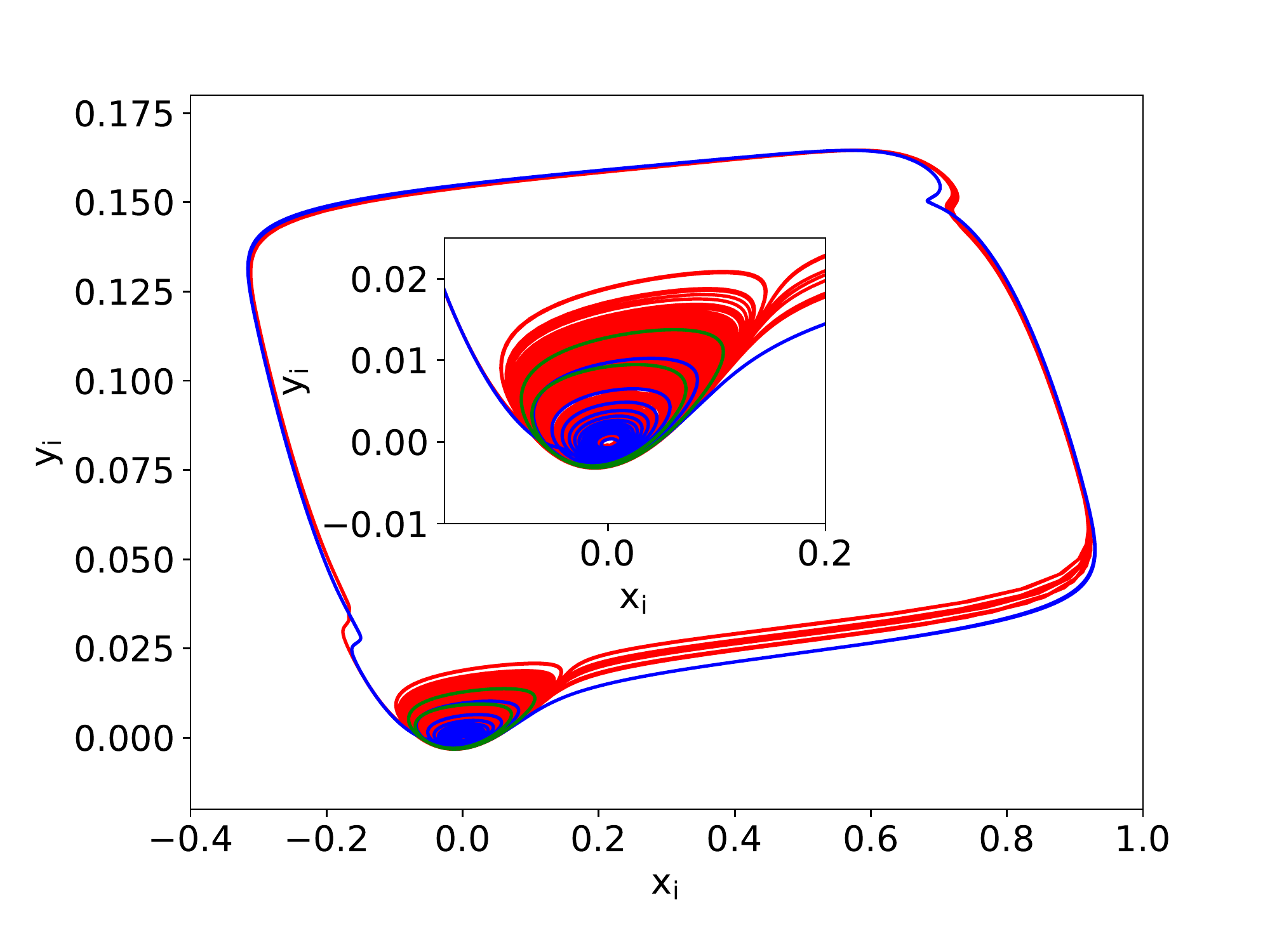}
\caption{Phase space representation of the various classes of attractors obtained on the synchronization manifold upon varying the coupling strength $M_2$. Small coupling $(M_2=0)$: limit cycle corresponding to mixed mode oscillations shown in blue; intermediate coupling $(M_2=0.00247)$: chaotic attractor corresponding to extreme events shown in red; large coupling $(M_2=0.0026)$: limit cycle corresponding to small amplitude oscillations shown in green. The inset shows the close-up view of the dynamics in the neighborhood of the origin. Other coupling parameters: $M_1=0.01$, $\tau_1=80$, $\tau_2=65$.}
\label{fig:Synch}
\end{figure}

\begin{figure*}
  \includegraphics[width=\linewidth]{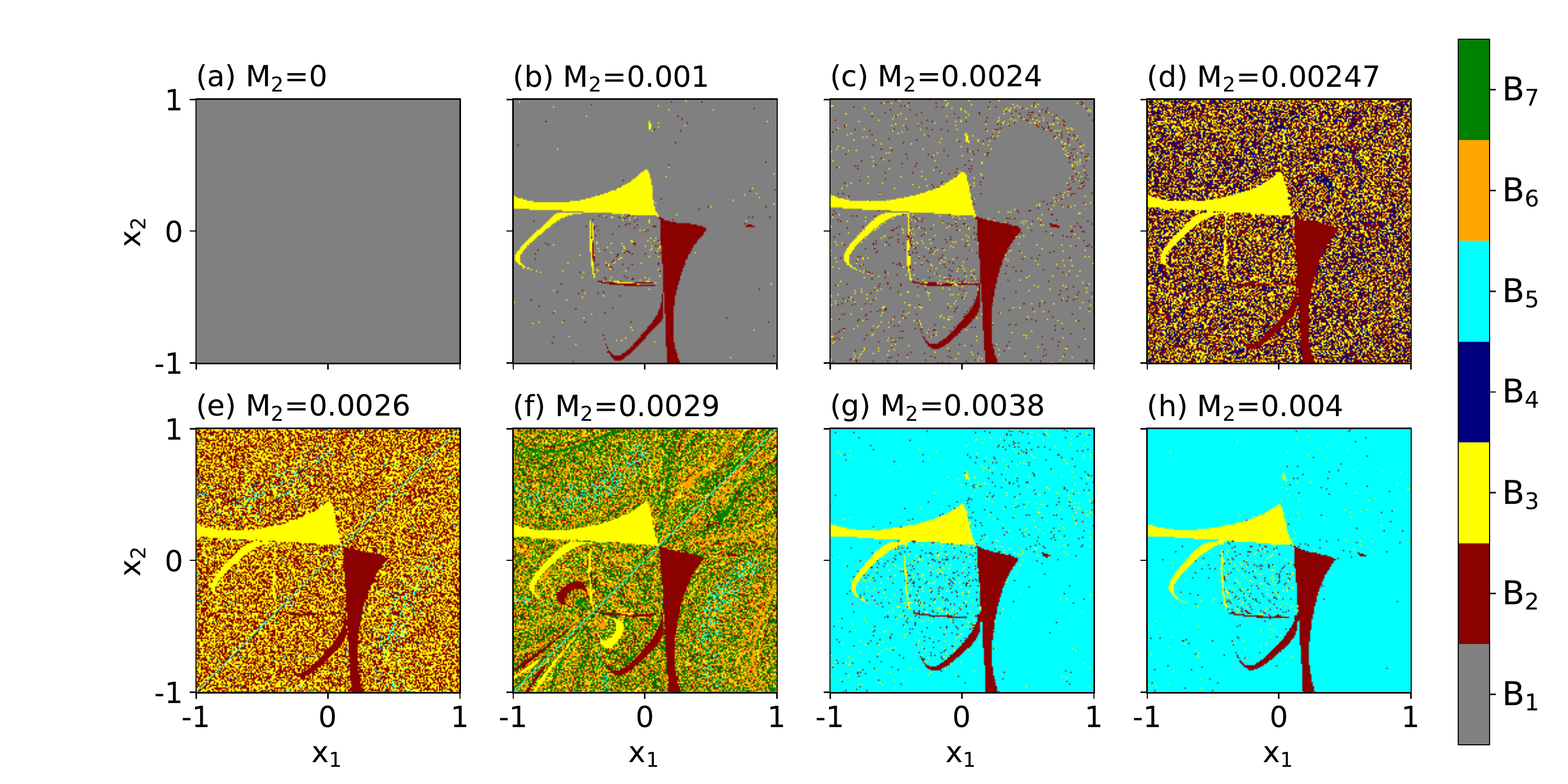}
\caption{Two dimensional slices of the phase space showing the basins of attraction at various values of coupling strength $M_2$. For each panel, the slice is taken at $y=y_1=y_2=0.01$. Other coupling parameters: $M_1=0.01$, $\tau_1=80$, $\tau_2=65$. The color code indicates the different basins of attraction \B1,$\ldots$,\B7 corresponding to the different attractors \A1,$\ldots$,\A7 (see text).}
\label{fig:Regimes}
\end{figure*}

\subsection{Regime 1}

For $M_2=0$, the total coupling strength $M=M_1+M_2=0.01$ is still less than the minimum coupling required for stabilization of the non-trivial fixed points (see Fig.~\ref{fig:FixedPoints}). The global attractor of the system, \A{1} is the limit cycle on the synchronization manifold corresponding to mixed mode oscillations (see Fig.~\ref{fig:Synch}: blue curve) and therefore, all initial conditions converge to it. This leads to the trivial basin structure \B{1} shown in Fig~\ref{fig:Regimes}a. The structure of \B{1} remains unchanged until $M_2 \approx 0.0003$ (or correspondingly $M \approx 0.0103$).

\subsection{Regime 2}

If $M_2$ is increased beyond $0.0003$, the reverse Hopf bifurcation stabilizes a pair of fixed points --- \A{2} and \A{3} --- placed symmetrically on either sides of the synchronization manifold and makes the system tri-stable. The two new attractors form tongue shaped basins --- \B2 and \B3 --- in the slice of phase space (see Fig.~\ref{fig:Regimes}b for an example). The trajectories starting within these tongue-shaped regions do not approach the synchronization manifold during the transient and converge to either of the fixed points directly. Note, that the diagonal line in the slice of phase space containing the basin of attraction shown represents the synchronization manifold of the system. Therefore, pairs of initial conditions which are symmetrically placed with respect to the diagonal either converge to the attractor on the synchronization manifold \A{1} (and belong to the basin \B{1}); or converge to the pair of fixed points \A{2} and \A{3} (and belong to basins \B{2} and \B{3} respectively) which themselves are symmetric with respect to the synchronization manifold. This makes the basins of attraction symmetric which is expected from the system in consideration. While the majority of the points belonging to \B{2} and \B{3} are contained in the tongue like structures, there are initial conditions which appear scattered in the area outside the tongues and yet converge to the fixed points \A{2} and \A{3}. As $M_2$ is increased further, the number of points belonging to \B{2} and \B{3} scattered outside the tongue-like structures increases (see Fig.~\ref{fig:Regimes}c).

\subsection{Regime 3}

\begin{figure}
  \includegraphics[width=0.95\linewidth]{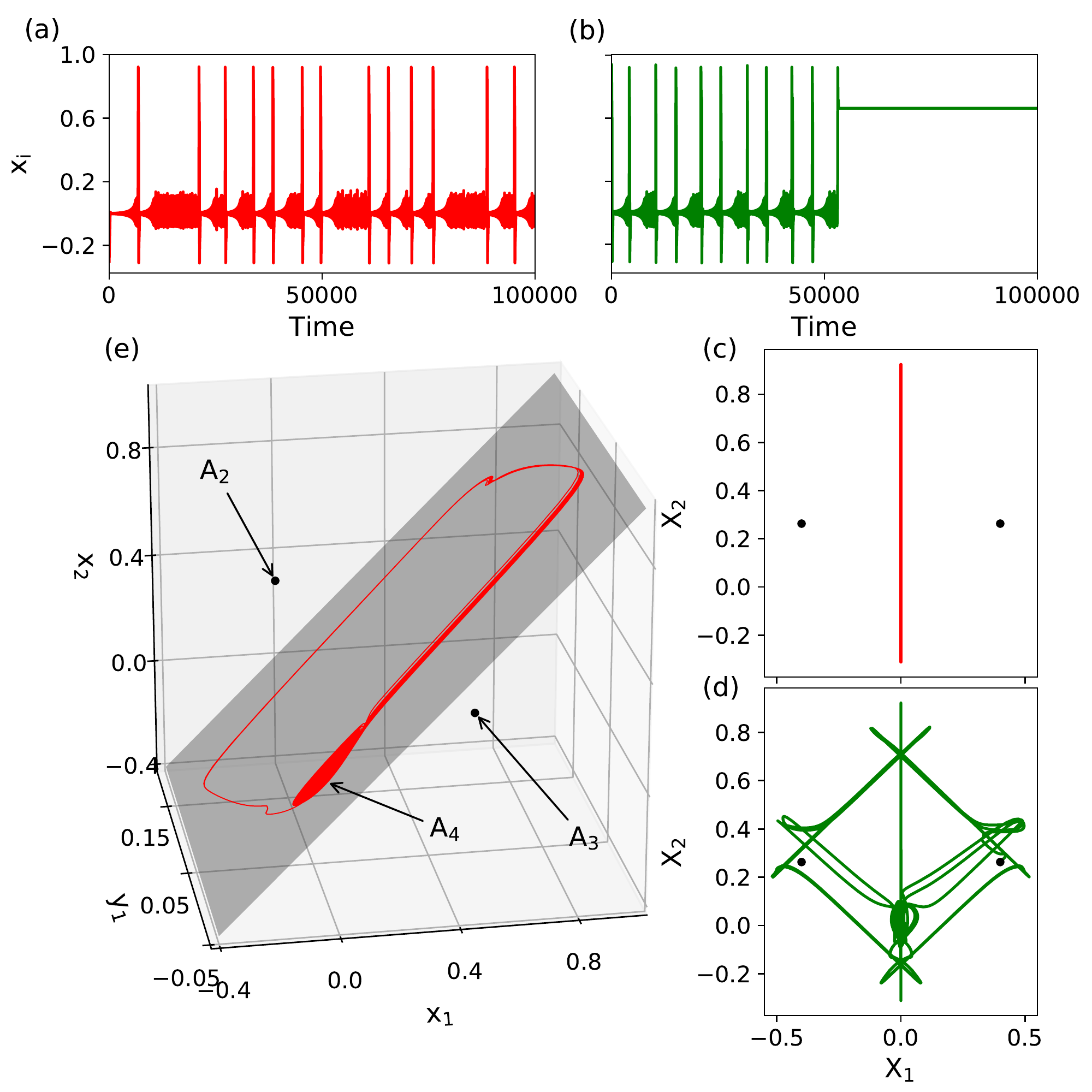}
  \caption{Various representations of the dynamics observed between $M_2 \approx 0.00245$ and $M_2 \approx 0.00255$. The attractors of the system are shown in a three dimensional projection of the phase space in (a). They include the red chaotic attractor \A1 on the invariant synchronization manifold shown in gray and the pair of stable fixed points \A2 and \A3. Time evolution of typical trajectories converging to attractors \A1 and \A3 are shown in (b) and (c) respectively. Another phase space representation of the trajectories in (b) and (c) are shown in (d) and (e) respectively where the transformed coordinates $\left( X_1, X_2 \right)$ are plotted. While the red trajectory converges quickly to the synchronization manifold, the green comes close to the manifold and diverges away from it multiple times before converging to the fixed point.}
\label{fig:Multistable}
\end{figure}

\begin{figure}
  \includegraphics[width=0.95\linewidth]{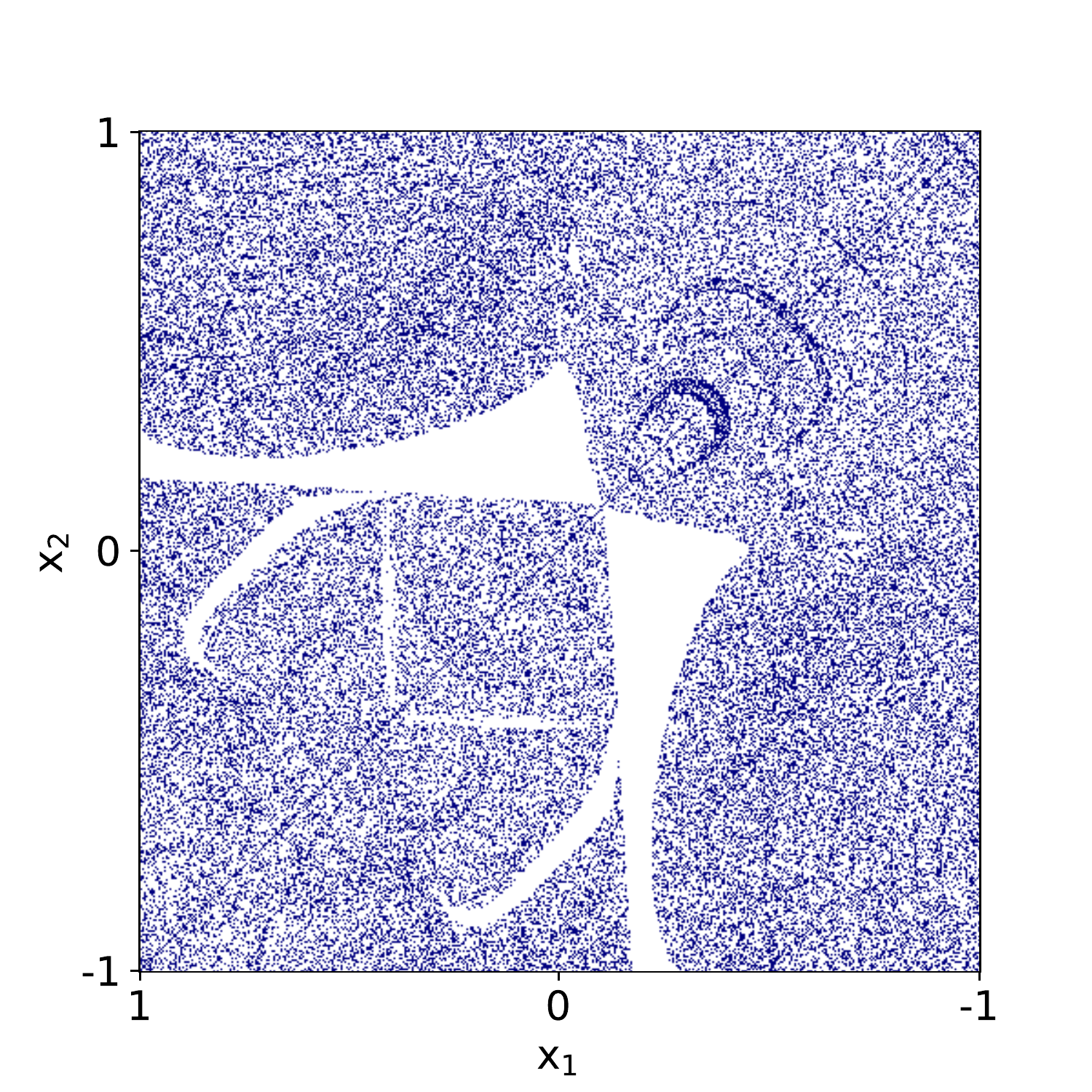}
  \caption{Basin of attraction corresponding to extreme events (navy blue) in Regime 3 as seen in a two-dimensional slice of the phase space. This is a plot of that basin of attraction only, compared to the plot of all basins of all coexisting attractors shown in Fig.~\ref{fig:Regimes}d.}
\label{fig:ExEv_Basin}
\end{figure}

So far, the trajectories on the synchronization manifold execute mixed mode oscillations comprising of several small amplitude oscillations followed by a single large amplitude oscillation or an event. While the number of small-amplitude oscillations between two consecutive events increases as $M_2$ is increased through Regimes 1 and 2, the overall dynamics on the synchronization manifold remains periodic in these regimes. Therefore, the inter-event-intervals throughout the long term trajectory remains a constant in time.

However, upon increasing $M_2$ beyond $0.00245$, we enter Regime 3 and the limit-cycle corresponding to the mixed mode oscillations undergoes a period-adding cascade to become a chaotic attractor \A{4} (see Fig.~\ref{fig:Synch}: red curve). In particular, the small amplitude oscillations between two successive events become highly chaotic, which results in an extremely high irregularity in the inter-event-intervals. These rare, recurrent and highly irregular events in such systems are known as extreme events and have been analyzed in detail in our recent work~\cite{PhysRevE.95.062219}.

The transition to Regime 3 is also accompanied by the emergence of an extremely rich structure of the basins of attraction. One of the distinct qualitative changes which occurs during the transition from Regime 2 to Regime 3 is the significant increase in the number of points which are in \B{2} and \B{3} but not in the tongue.  As can be clearly seen in Fig.~\ref{fig:Regimes}d, the phase space seems now to be comprised of two distinct types of regions: The `pure' regions where neighboring points belong to only one particular basin of attraction; and the `mixed' regions where neighboring points may belong to any of the three basins of attraction. Notably, the pure regions seem to contain points only belonging to \B{2} or \B{3} and not to \B{4}. In other words, the entire basin \B{4} seems to be contained in the mixed regions of the phase space, which is illustrated by plotting only the points of \B{4} in Fig.~\ref{fig:ExEv_Basin}. The quantitative aspects of the characteristics of these regions will be discussed in more detail in the following section.

The emergence of those two distinct regions in the phase space --- denoted as `pure' and `mixed' --- also impacts the transients of the trajectories which do not start on the synchronization manifold. A trajectory which starts in one of the pure regions converges to the corresponding fixed point without repeatedly approaching the neighborhood of the synchronization manifold, hence yielding a relatively short transient. The trajectories starting in the mixed regions, however, may approach the neighborhood of the synchronization manifold many times and trace out the close proximity of the chaotic attractor on the synchronization manifold before being ejected and converge finally to one of the stable fixed points. This leads to possibly very long transients where the dynamics of the trajectory --- which will eventually converge to a fixed point --- resembles closely the extreme event dynamics of the trajectory which has converged to the chaotic attractor on the synchronization manifold (see Fig.~\ref{fig:Multistable}).

\subsection{Regime 4}

On increasing the coupling strength beyond $M_2 \approx 0.00255$, the attractor on the synchronization manifold changes from being the chaotic set \A{4} to a small-amplitude limit cycle \A{5} (see Fig.~\ref{fig:Synch}: green curve). This changes the long-term motion for trajectories starting on the synchronization manifold from being comprised of small chaotic oscillations interspersed by irregularly appearing large events to a periodic oscillation with small amplitude. Initial conditions not starting on the synchronization manifold may still converge either to the non-trivial fixed points \A{2} and \A{3}; or the attractor on the synchronization manifold \A{5}. The structure of the basins of attraction for such a regime is shown in Fig.~\ref{fig:Regimes}e. Note that the phase space still has the mixed regions which now mostly consist of points from the basins \B{2} or \B{3}. Nevertheless, points belonging to \B{5} can still be found in parts of the mixed regions on either side of the diagonal. This correlates with the numerical observation that although a large majority of the initial conditions starting away from the synchronization manifold converge to either \A{2} or \A{3}; there are initial conditions not on the synchronization manifold which converge to \A{5}. Note that as the coupling strength $M_2$ is increased, the periodicity of the limit-cycle decreases due to a reverse period-doubling cascade.

\subsection{Regime 5}

\begin{figure}
  \includegraphics[width=0.95\linewidth]{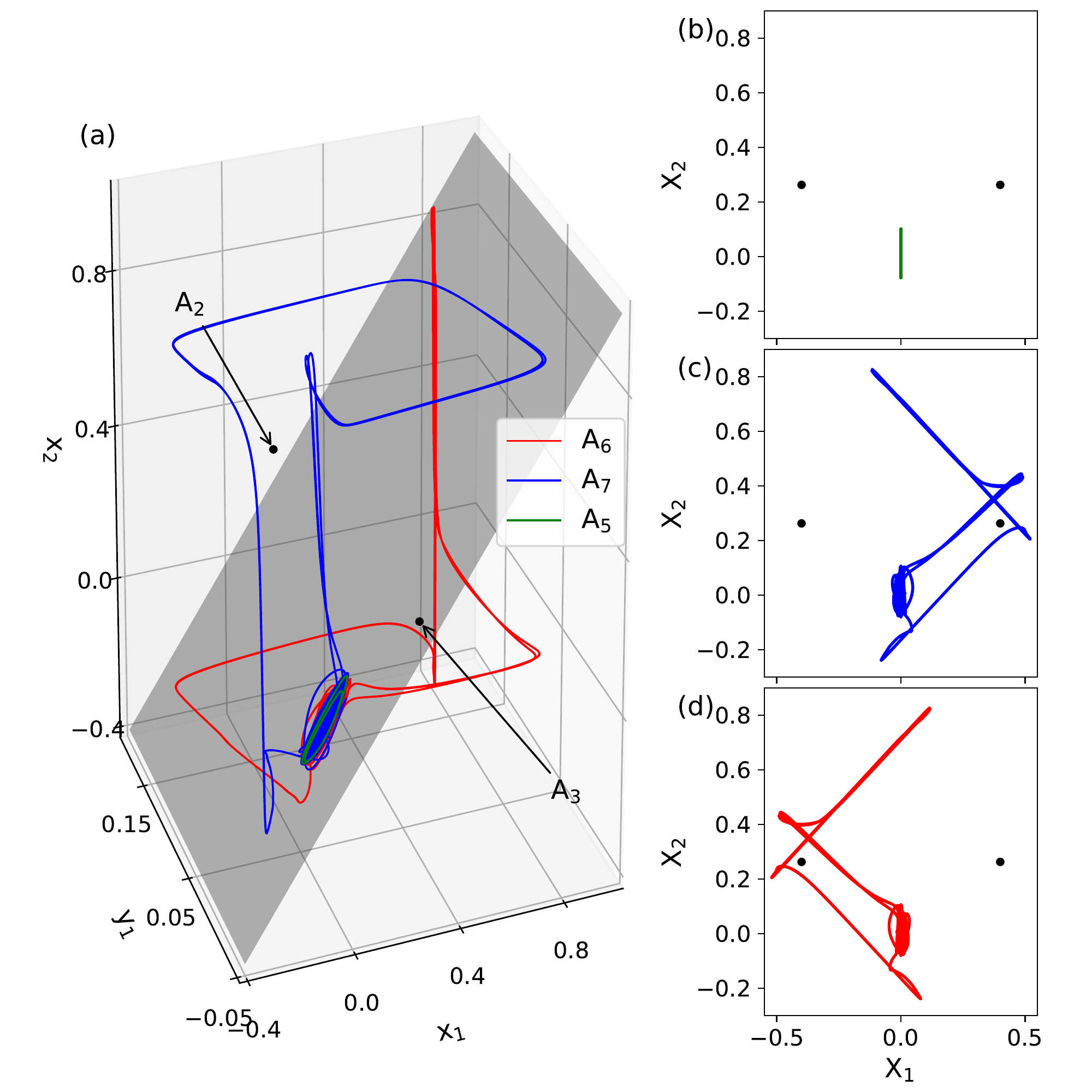}
  \caption{Various representations of the dynamics of the system observed between $M_2 \approx 0.0026$. The attractors of the system are shown in a three dimensional projection of the phase space in (a). They include the green limit cycle \A1 on the invariant synchronization manifold shown in gray, the pair of stable fixed points \A2 and \A3; and the blue and red non-synchronized chaotic attractors \A4 and \A5. Phase space representations of trajectories on attractors \A1, \A4 and \A5 in the transformed co-ordinates $\left( X_1, X_2 \right)$ are plotted in (b), (c) and (d), respectively.}
\label{fig:VeryMultistable}
\end{figure}

If the coupling strength is increased beyond $M_2 \approx 0.00285$, an additional pair of chaotic attractors, \A{6} and \A{7}, appear on either sides of the synchronization manifold. The system therefore, now contains a total of 5 co-existing attractors (see Fig.~\ref{fig:VeryMultistable}): a small amplitude limit-cycle \A{5}, a pair of non-trivial fixed points \A{2} and \A{3}, and a pair of non-synchronized chaotic attractors \A{6} and \A{7}.

A trajectory that converges to \A{6} or \A{7} executes nearly synchronous small-amplitude oscillations interspersed by single highly asynchronous large amplitude oscillations. In phase space, the trajectory remains extremely close to the synchronization manifold during the small-amplitude oscillations and diverges away from it during the large-amplitude oscillation. Note that, although the trajectory exhibits a dynamics similar to the attractor containing extreme events, i.e. it comprises of many small amplitude oscillations followed by a large amplitude oscillation, the dynamics in this case cannot be classified as extreme events as the events are not irregular and not rare enough.

The emergence of \A{6} and \A{7} leads to an additional richness in the structure of the basins of attraction (see Fig.\ref{fig:Regimes}f). Similar to the previous two regimes, the phase space appears to be partitioned into mixed and pure regions. However, each of the pure regions in this regime can also belong to either of the chaotic attractors \A{6} or \A{7} in addition to the previously present stable fixed points. The mixed regions on the other hand, contains points belonging to the basins of all attractors in the system including the small-amplitude limit cycle \A{5} on the synchronization manifold.

\subsection{Regime 6}

Beyond $M_2 \approx 0.00360$, the chaotic invariant sets outside the synchronization manifold are no longer stable and the only attractors which remain in the system are the fixed points \A{2} and \A{3}, and the small-amplitude limit cycle \A{5}. This is qualitatively similar to Regime~2 with the attractor on the synchronization manifold being the limit-cycle corresponding to small-amplitude instead of mixed mode oscillation.

This similarity in the nature of attractors is also reflected in the basin structure (see Fig.~\ref{fig:Regimes}g). The basins of the fixed points comprise mostly of the tongue like structures emanating from the synchronization manifold, and additional isolated points scattered elsewhere in phase space. The rest of the phase space forms the basin of attraction of the limit cycle \A{5}. Note that there are no `mixed' regions in phase space anymore. Moreover, the number of points which belong to the basins \B{2} and \B{3} and yet do not belong to the tongue-like structures decrease as the coupling strength is increased up to $M_2 \approx 0.0042$ (see Fig.~\ref{fig:Regimes}h).

On increasing $M_2$ even further, the system exhibits more interesting dynamics including small-amplitude chaotic oscillations and stabilization of the origin. However, the detailed analysis the system in these regimes is beyond the scope of this paper.

\begin{figure*}
  \includegraphics[width=\linewidth]{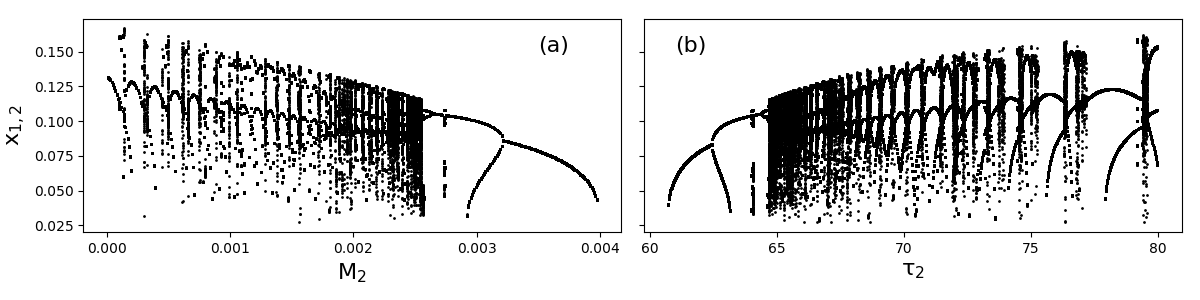}
  \caption{Bifurcation diagrams showing the Poincar\'{e} section (obtained by fixing $x_1=x_2$ and $y_1=y_2=0.01$) of the trajectories on the synchronization manifold for varying coupling parameters $M_2$ and $\tau_2$. Parameters: (a) $\tau_2=65$, (b) $M_2=0.00245$. Common Parameters: $M_1=0.01$, $\tau_1=80$.}
\label{fig:Bifurcation}
\end{figure*}

The changes in dynamics on the synchronization manifold presented in this section can be summarized by the bifurcation diagram in Fig.~\ref{fig:Bifurcation}. As the coupling strength increases, the stable limit-cycle corresponding to the mixed mode oscillations, \A1,  on the synchronization manifold undergoes a period-adding cascade to eventually become a chaotic attractor corresponding to an attractor \A4 containing extreme events. This chaotic attractor looses stability and a high periodicity small-amplitude limit cycle, \A5 emerges which thereafter undergoes a reverse period-doubling cascade to finally become a period-one limit-cycle.

The bifurcation diagram for varying time-delay $\tau_2$ with fixed coupling strength $M_2$ is also plotted in Fig.~\ref{fig:Bifurcation}. The latter is qualitatively similar to the first one except for the direction of changes in the dynamics. In other words, the qualitative changes observed in the system as coupling strength is increased follow the same order as the qualitative changes observed as the time-delay is decreased. This implies that the dynamical regimes and the corresponding basin structures described in this section can be obtained by varying either of the coupling parameters or even a combination of both.

\section{Characteristics of Basins of Attraction}
\label{sec:Characteristics}

In the last section, we noted that the basin structure in Regimes 3, 4 and 5 partitions in the phase space into pure and mixed regions. Here, we demonstrate that in these regimes the basins of attraction of certain attractors possess a riddled structure and hence, are fundamentally different from the Regimes 1, 2 and 6 where mixed regions do not exist in the phase space. We also highlight that such a property can have crucial consequences for the dynamics particularly when the occurrence of extreme events is involved.

In order to show that the basins in Regimes 3, 4 and 5 are riddled, we first compare the structure of basin boundaries in Regimes 3, 4 and 5 with those of Regimes 2 and 6. We thereby emphasize that in Regimes 3, 4 and 5, there are regions in phase space where an arbitrarily small perturbation in the initial conditions can lead to a trajectory converging to a different attractor.

\begin{figure*}
  \includegraphics[width=0.9\linewidth]{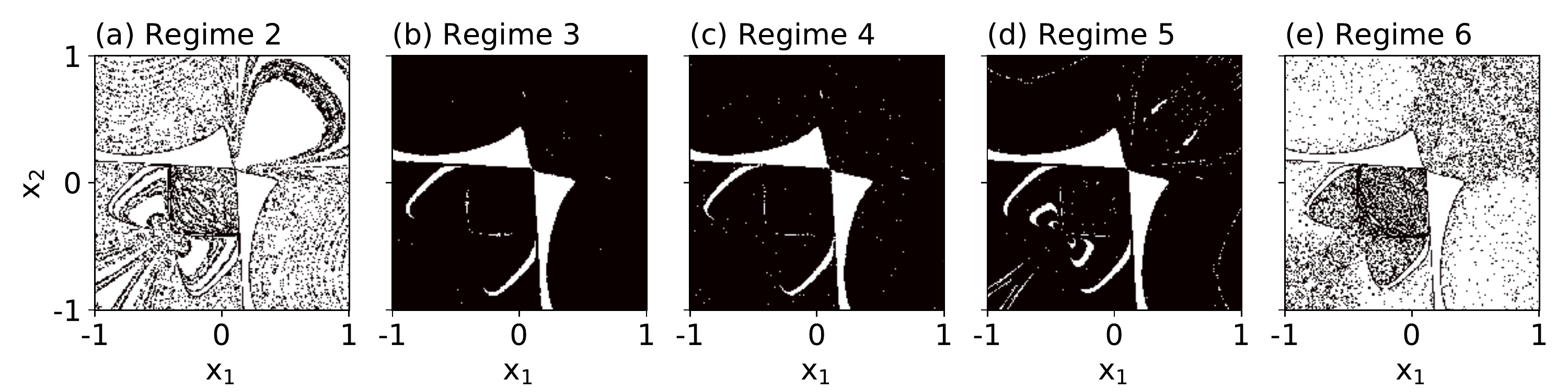}
  \caption{Two dimensional slices of the phase space at $y=y_1=y_2=0.01$, showing interior (shown in white) and boundary points (shown in black) of the basins of attraction in various dynamical regimes. For the particular Regimes 2, 3, 4, 5 and 6, the couplings strengths used are $M_2=0.0024$, $0.00247$, $0.0026$, $0.0029$ and $0.0038$, respectively (as in Fig.~\ref{fig:Regimes}).}
\label{fig:Pure_Mixed}
\end{figure*}

We start our analysis by assigning all points in phase space to two categories with regards to their position in their respective basins of attraction: interior points and boundary points. A point is said to be an interior point if all the points in its infinitesimal neighborhood belong to the same basin of attraction as the point under consideration. All other points are classified as boundary points. While the exact classification of points is not possible in numerical computations since we always deal with a certain resolution; we approximate this classification by constructing a mesh of $512 \times 512$ points spanning the two dimensional slice of the phase space. Thereafter, we assume that the next neighbors of each point in the mesh belong to its infinitesimal neighborhood. The results obtained are presented in Fig.~\ref{fig:Pure_Mixed}. The accuracy of the method can be increased by starting with a finer mesh. However, it was verified using a $1024 \times 1024$ grid that the results obtained are qualitatively identical to the results presented here.

Let us first analyze the phase space in Regimes 2 and 6 (see Fig.~\ref{fig:Regimes}). Most of the phase space is comprised of continuous two dimensional regions belonging to a particular basin of attraction. However, the phase space also contains numerous isolated points, each of which, belongs to a particular basin of attraction, say of attractor \A{i}, but is surrounded completely by points belonging to a different basin of attraction, say of attractor \A{j}. Note that all the points surrounding the isolated point belong to the basin of attraction of the same attractor \A{j}. However, that attractor \A{j} is different from the attractor \A{i} corresponding to the isolated point. From Fig.~\ref{fig:Pure_Mixed}, it can be seen that for Regimes 2 and 6, boundary points (colored black) form one dimensional curves separating regions of interior points (colored white) belonging to different basins of attraction. Boundary points also mark the isolated points and their immediate neighbors.

\begin{figure}
  \includegraphics[width=\linewidth]{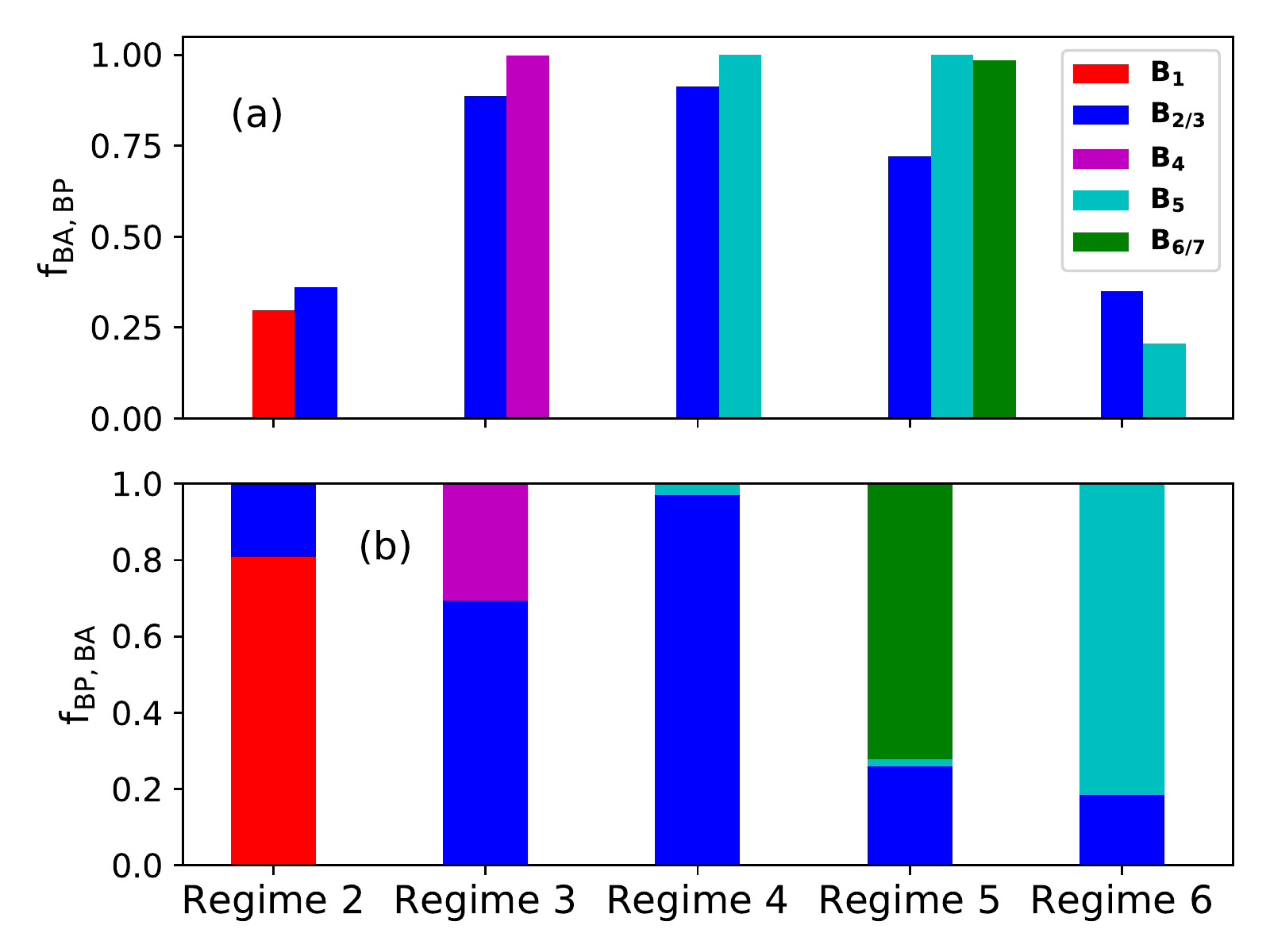}
  \caption{Characteristics of boundary points of the basins of attraction in various regimes. Fractions of points in the various basins of attraction $f_{BA,BP}$ which are boundary points are shown in (a) as bar plots. Whereas (b) shows a stacked histogram plot depicting the composition of the set of all boundary points in terms of basins of attraction to which each of the boundary points belongs. The slices of basins of attraction used for the analysis are taken from Fig.~\ref{fig:Regimes}.}
\label{fig:Bar_Plot}
\end{figure}

The situation is evidently different in Regimes 3, 4 and 5 where boundary points seem to fill up two dimensional regions in phase space which are colored in black. Notably, the boundary points appear to cover the mixed regions of the phase space entirely whereas the interior points cover the pure regions. This implies that every point in the mixed region has at least one point in its immediate neighborhood that belongs to a different basin of attraction than itself. Therefore, for any trajectory starting from the mixed region which converges to a particular attractor, there exists an infinitesimally small perturbation to that initial condition which would push the trajectory across a basin boundary and cause it to converge to a different attractor. Plotting the fraction of boundary points in each basin of attraction in Regimes 2 through 6 (see Fig.~\ref{fig:Bar_Plot}a), it can be inferred that the basins of attraction of attractor \A4 in Regime 3, and attractor \A5 in Regimes 4 and 5 are completely contained in the mixed regions of the phase space as they are entirely composed of boundary points. This indicates that attractor \A4 in Regime 3 and \A5 in Regimes 4 and 5 have riddled basins of attraction as each point belonging to the basins of attraction of these attractors has in its immediate neighborhood, a point belonging to the basin of attraction of another attractor.

The significance of a riddled basin in Regime 3 is greatly increased as the attractor possessing the riddled basin corresponds to the occurrence of extreme events. The basin in consideration, \B4, is riddled in basins \B2 and \B3 which correspond to the fixed point attractors. Note also, that an initial condition in \B2 or \B3 which belongs to the mixed region of the phase space exhibits a long transient during which it closely traces the chaotic attractor corresponding to the occurrence of extreme events before converging to the fixed points. This underlines an important property of the system under consideration: (a) any initial condition in the mixed region of the phase space can potentially exhibit extreme events for a long time, if not perpetually; and (b) due to the riddled nature of the basins of attraction, even a very small perturbation in initial conditions in the mixed region can change a system from exhibiting extreme events as a transient behavior to exhibiting extreme events forever.

Although a stable chaotic attractor corresponding to extreme event generation does not exist for Regimes 4 and 5, having a riddled basin structure in these regimes in still important as the long transients which closely resemble extreme events may be observed for trajectories starting from the mixed regions of the phase space. Additionally we note that, Regime 5 contains two stable chaotic attractors \A6 and \A7. While the events exhibited by the trajectories converging to these attractors are not irregular enough to be classified as extreme events, they are still recurrent and have a significantly larger amplitude than the typical oscillation of the system which may considerably affect the system. In order to illustrate this point, we plot in Fig.~\ref{fig:Bar_Plot}b, stacked histograms showing the fraction of boundary points which belong to the respective attractors. From Fig.~\ref{fig:Bar_Plot}a and Fig.\ref{fig:Bar_Plot}b, it can be inferred that not only a major fraction of \B6 and \B7 belongs to the mixed regions of phase space, but also that the points belonging to \B6 and \B7 constitute a large fraction of all points in the mixed region. This implies: (a) if behavior exhibiting a regular occurrence of events is the desired state of the system, the choice of plausible initial conditions is restricted to the small pure regions in \B6 and \B7; as other initial conditions in \B6 or \B7 are in the mixed region and hence, are vulnerable to small perturbations and (b) an initial condition in the mixed region of the phase space is most likely to result in regular behavior containing frequently occurring events.

\begin{figure*}
  \includegraphics[width=0.9\linewidth]{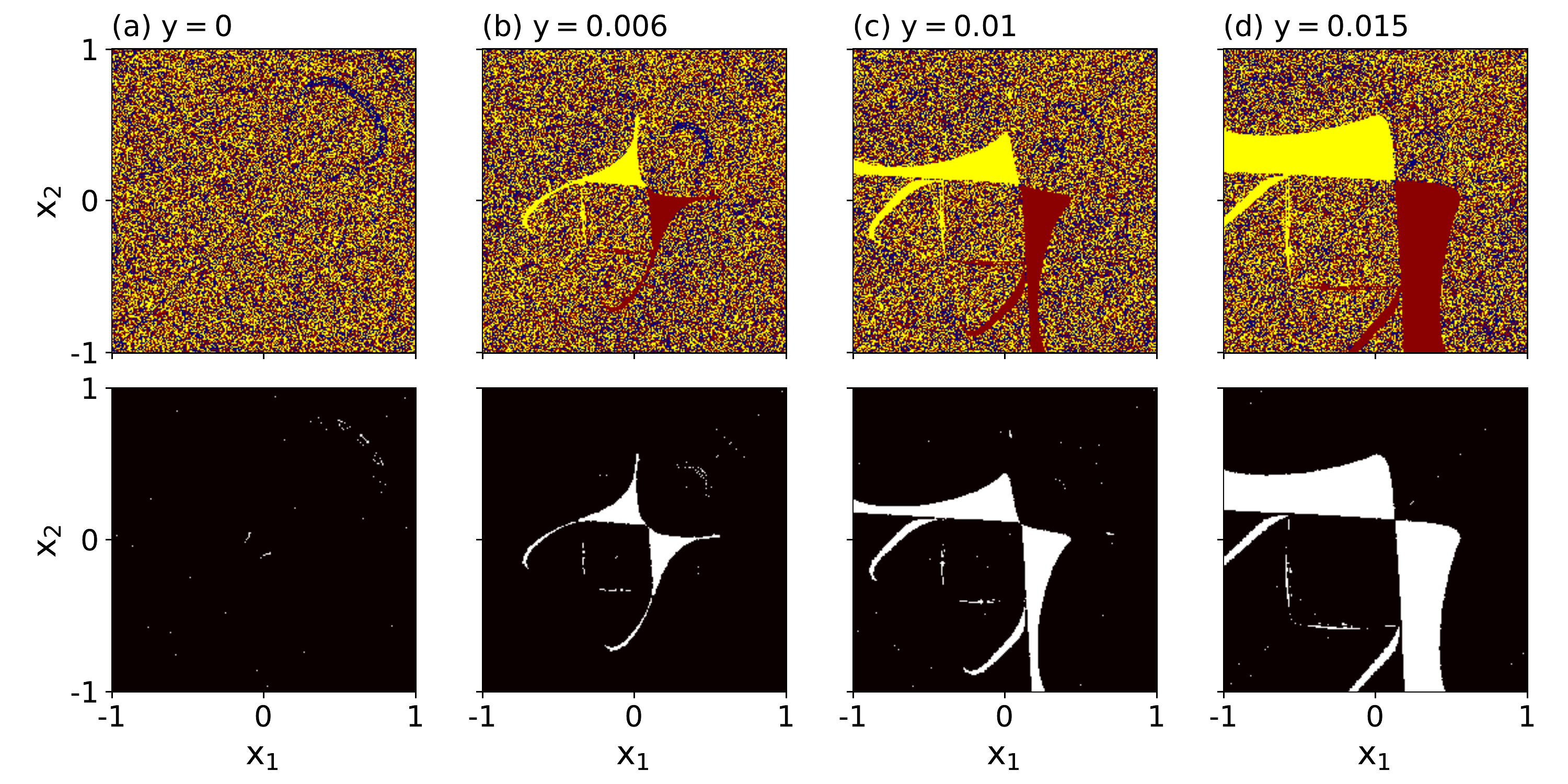}
  \caption{The top row shows two dimensional slices of phase space, color coded for the different basins of attraction in the system, for varying values of $y=y_1=y_2$ when coupling strength is fixed at $M_2=0.00247$ (as in Fig.~\ref{fig:Regimes}). The bottom row shows the classification of all points as interior or boundary points. The color code for the various basins of attraction is the same as in Fig.~\ref{fig:Regimes} and Fig.~\ref{fig:Pure_Mixed}.}
\label{fig:Slices}
\end{figure*}

In order to ensure that the observations regarding the riddled nature of basins of attraction is not a manifestation of the specific choice of the slice in phase space used for obtaining the basins of attraction, we present plots of basins of attraction in other slices in Fig.~\ref{fig:Slices}. These slices are obtained by choosing various values of $y=y_1=y_2$ for $M_2=0.00247$ (Regime 3). From the figure, we observe that although the size and shape of the tongues change as $y$ is varied, the qualitative structure of the basin remains consistent. Again we observe a partitioning of the phase space into pure and mixed regions and the basin of attraction of the attractor \A3 corresponding to extreme events is completely contained in the mixed regions. A similar analysis of the phase space in other regimes also reveal results which are in agreement with those presented previously in this section.

\begin{figure}
  \includegraphics[width=\linewidth]{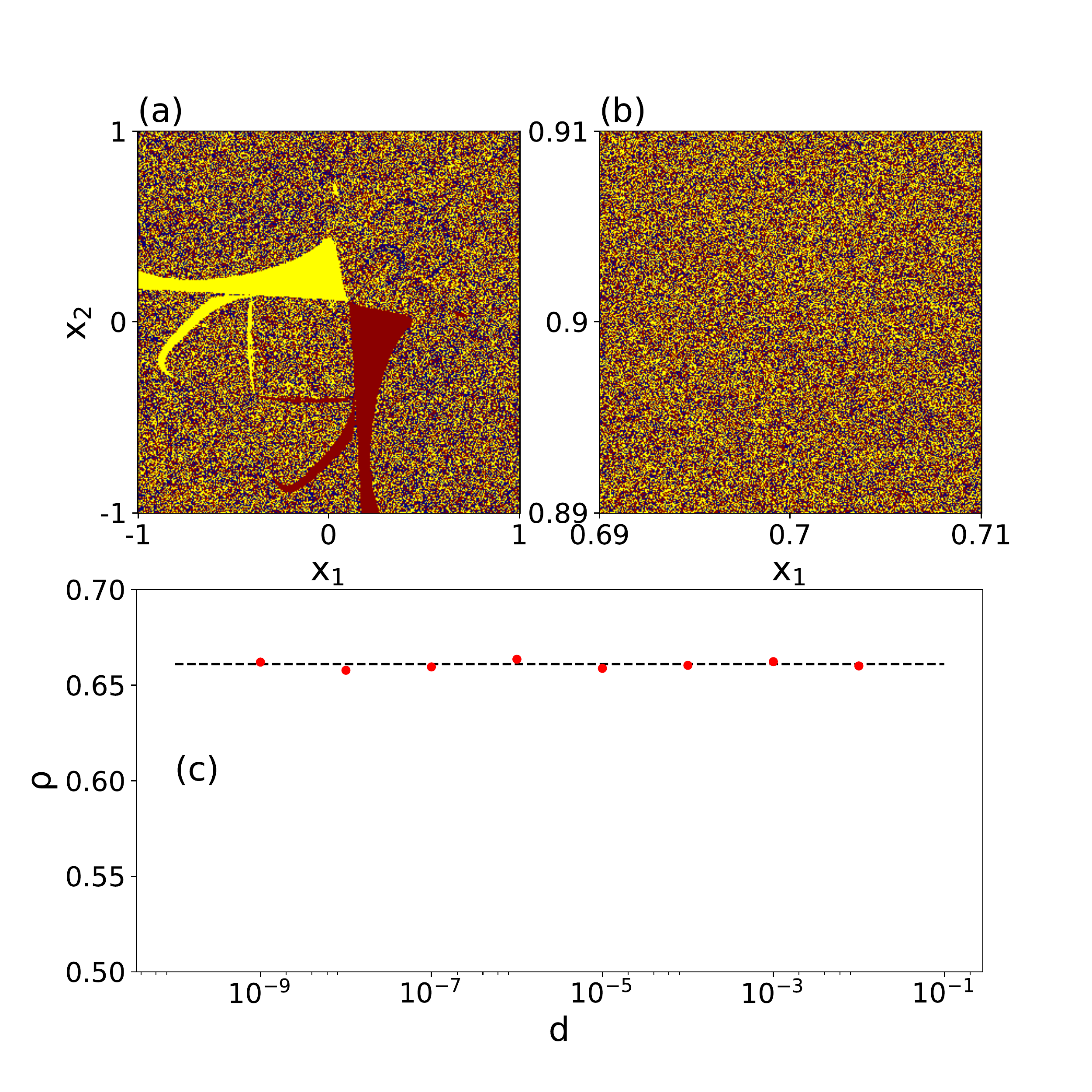}
  \caption{Basins of attraction in the two dimensional projection of the phase space shown in (a) and the zoomed in view of a portion of its mixed region in shown in (b). Color code for (a) and (b) is the same as in Fig.~\ref{fig:Regimes}. The uncertainty exponent $\rho$ for the region in (b) is plotted against the distance between points $d$ in (c) using the solid red dots. The dashed line gives the best fit for the points in red.}
\label{fig:Pair_Invariance}
\end{figure}

A very well known method to compute the dimensions of a fractal basin boundary is the computation of the uncertainty exponent $\alpha$~\cite{GREBOGI1983415}. In order to compute the exponent $\alpha$ in our system, we choose a part of the mixed region in the two dimensional slice of the phase space (for the zoomed in version of the selected region with a resolution of $10^{-12}$, see Fig.~\ref{fig:Pair_Invariance}(b)). We then choose a distance, $\varepsilon$ and randomly select 1000 pairs of initial conditions from the region such that the distance between each pair is $\varepsilon$. For varying values of $\varepsilon$, we then plot the fraction of pairs of initial conditions, $f(\varepsilon)$ such that each initial condition in the pair converges to a different attractor. The expected relation between $f(\varepsilon)$ and $\varepsilon$ for a fractal basin boundary is
\begin{equation}
  f(\varepsilon) \sim \varepsilon^\alpha,
  \label{eq:Uncertainty}
\end{equation}
where the uncertainty exponent $\alpha$ is the difference between the dimension of the state space and the dimension of the basin boundary. Though this method of final state sensitivity has been developed only for systems defined in a finite dimensional phase space, we believe that it also provides similar insights into the basin structure of systems in an infinite dimensional phase space. Our analysis shows that for the system in consideration, $\alpha = 7.476 \times 10^{-7}$ for $M_2=0.00247$ (see Fig.~\ref{fig:Pair_Invariance}). This value is very close to zero implying that the dimensions of the basin boundary is approximately equal to the dimension of the state space. This is in accordance with the results shown in Fig.~\ref{fig:Pure_Mixed} where the boundary points seemed to span a two dimensional region in the two dimensional slice of the phase space. Our results also agree with previous studies of riddled basins of attraction where the uncertainty exponent $\alpha$ has been reported to be approximately zero~\cite{doi:10.1063/1.4954022}. Although the coupling strength chosen for Fig.~\ref{fig:Pair_Invariance} is in Regime 3, the results for Regimes 4 and 5 where riddled basins are also observed are similar to the ones presented here.

\section{Conclusions}
\label{sec:Conclusions}

In this study, we have explored in detail the various regimes of multistability and the structure of the corresponding basins of attraction exhibited by a system of two identical FitzHugh-Nagumo units connected to each other using two coupling delays. In our analysis, we have focused on a parameter interval which includes the regime where this system exhibits extreme events. Depending on the coupling strength we obtain up to 5 different co-existing attractors. Due to the symmetry of the system, one of the attractors is located on the synchronization manifold, while the other attractors lie outside this manifold. We find that the basin structure of the system becomes progressively rich and complex as we approach the parameter regime where extreme events are observed. While many basins of attraction are fractal, we also find basins of attraction which are riddled. The significance of this result is increased as one of the riddled basins corresponds to the extreme event dynamics. To classify these basins as riddled, we compute the uncertainty exponent, which is found to be very close to zero giving a strong indication of a riddled basin. Although riddled basins have been reported previously in many systems, our investigation is, to the best of our knowledge, the first evidence of a riddled basin in an infinite dimensional system such as a delay-coupled system. Additionally, we have shown that the method of final state sensitivity which was originally developed for finite dimensional systems can be successfully employed in the case of infinite-dimensional systems, where the computation of basins of attraction is particularly difficult.

Similar to the findings of previous studies, we show that in case of a riddled basin the phase space can be divided into pure and mixed regions. A crucial aspect of our analysis is that one of the basins which shows riddling belongs to an attractor which contains extreme events. This basin of attraction is completely confined to the mixed regions of the phase space. This has an important consequence for the overall dynamics: While any trajectory starting from the pure regions in phase space leads to a safe dynamics far away from extreme events, the trajectories starting in the mixed regions of phase space may or may not converge to the state containing extreme events. Those initial conditions in the mixed region are extremely sensitive with respect to perturbations. Already very tiny perturbations would be sufficient to push the trajectory to a dynamics which contains extreme events. Therefore, we obtain a high risk of ending up in a state of extreme events and which of the initial conditions lead to them is not predictable.

\section*{Acknowledgments}

The authors would like to thank G. Ansmann, A. Choudhary, P. H\"ovel, E. Knobloch, K. Lehnertz, C. Masoller, E. Sch\"oll, S. Wieczorek and J. A. Yorke for fruitful discussions and critical suggestions. This work was supported by the Volkswagen Foundation (Grants No.~88459). The simulations were performed at the HPC Cluster CARL, located at the University of Oldenburg (Germany) and funded by the DFG through its Major Research Instrumentation Program (INST 188/157-1 FUGG) and the Ministry of Science and Culture (MWK) of the State Lower Saxony.

\bibliography{Ref}

\end{document}